\newcommand{\CLASSINPUTtoptextmargin}{0.9 in}
\newcommand{\CLASSINPUTbottomtextmargin}{1.1 in}
\documentclass[conference, 10pt]{IEEEtran}
\IEEEoverridecommandlockouts
\usepackage{stmaryrd}
\usepackage{amsmath}
\usepackage{amssymb}
\usepackage{psfrag}
\usepackage{graphicx}
\usepackage{epstopdf}
\usepackage{multirow}
\usepackage{booktabs}
\usepackage{siunitx}
\usepackage{cite}
\usepackage{algorithmic}
\usepackage{algorithm}
\usepackage{svg}
\usepackage{mathtools}
\usepackage{dutchcal}
\usepackage{hyperref}
\usepackage{bm}

\DeclareMathOperator*{\argmin}{argmin}

\usepackage{float}
\ifCLASSOPTIONcompsoc
\usepackage[caption=false,font=normalsize,labelfont=sf,textfont=sf]{subfig}
\else
\usepackage[caption=false,font=footnotesize]{subfig}
\fi

\usepackage{flushend}

\begin{document}

\title{
Neural Network-based Two-Dimensional Filtering for OTFS Symbol Detection
}
\author{Jiarui Xu, Karim Said, Lizhong Zheng, and Lingjia Liu
\thanks{J. Xu, K. Said, and L. Liu are with Wireless@Virginia Tech, the Bradley Dept. of ECE at Virginia Tech. L. Zheng is with the EECS Department at the Massachusetts Institute of Technology. 
The work of J. Xu, K. Said, L. Zheng, and L. Liu was supported by the National Science Foundation under Grant NSF/CCF-2003059. 
}%
}
%



\maketitle

\begin{abstract}
Orthogonal time frequency space (OTFS) is a promising modulation scheme for wireless communication in high-mobility scenarios.
Recently, a reservoir computing (RC) based approach has been introduced for online subframe-based symbol detection in the OTFS system, where only the limited over-the-air (OTA) pilot symbols are utilized for training.
However, the previous RC-based approach does not design the RC architecture based on the properties of the OTFS system to fully unlock the potential of RC.
This paper introduces a novel two-dimensional RC (2D-RC) approach for online symbol detection on a subframe basis in the OTFS system.
The 2D-RC is designed to have a two-dimensional (2D) filtering structure to equalize the 2D circular channel effect in the delay-Doppler (DD) domain of the OTFS system.
With the introduced architecture, the 2D-RC can operate in the DD domain with only a single neural network, unlike our previous work which requires multiple RCs to track channel variations in the time domain.
Experimental results demonstrate the advantages of the 2D-RC approach over the previous RC-based approach and the compared model-based methods across different modulation orders.

\end{abstract}


%
\IEEEpeerreviewmaketitle


\section{Introduction}

Next-generation wireless communication systems are expected to support reliable communication quality in high-mobility scenarios, such as high-speed railways, unmanned aerial vehicles, and low earth orbit~\cite{series2015imt}.
Orthogonal time frequency space (OTFS) modulation~\cite{hadani2017orthogonal} has emerged as a potential technology to combat the high Doppler effect in such scenarios.
Different from orthogonal frequency division multiplexing (OFDM) which multiplexes information symbols in the time-frequency (TF) domain, OTFS is a two-dimensional (2D) modulation scheme that transmits information symbols in the delay-Doppler (DD) domain.
In the DD domain, the channel has a sparse and compact representation~\cite{hadani2017orthogonal}, making it intrinsically suitable for reliable data transmission in time-varying channels.

The benefits of adopting OTFS modulation in high-mobility scenarios have attracted substantial interest in investigating equalization techniques for the OTFS system.
Existing approaches can be mainly categorized into two classes: model-based methods and learning-based approaches.
Model-based approaches are designed based on analyzing the input-output relationship and the structure of the equivalent channel matrix in the OTFS system.
Specifically, linear minimum mean square error (LMMSE) based linear equalizers, such as the work in~\cite{zou2021low}, are introduced to conduct low-complexity detection by taking advantage of the channel structure.
To achieve better performances, non-linear detectors, such as the message passing algorithm (MPA) based approaches~\cite{raviteja2018interference, liu2021message, zhang2021low}, are developed to approximate the maximum a \emph{a posteriori} (MAP) performance with low complexity by leveraging the channel sparsity.
While model-based approaches are explainable and easy to analyze, the assumptions of simplified system models and perfect channel knowledge may degrade the performance when such assumptions do not hold.

Learning-based detection approaches offer a promising solution for OTFS symbol detection in high-mobility scenarios by leveraging the power of neural networks (NNs).
For example, in~\cite{enku2021two}, a two-dimensional convolutional neural network (2D CNN) method is developed which adopts MPA for data augmentation and then processes the received signal as a 2D image with the CNN.
However, such approaches require extensive training samples along with a long training time.
While recent advances, such as GAMP-NET~\cite{zhang2022gaussian}, have attempted to unfold existing iterative model-based algorithms with NNs to reduce the number of trainable parameters, they still have a high demand for the amount of training data. 
Moreover, deep unfolding approaches also rely on perfect knowledge of channel state information (CSI), limiting their usage in practical scenarios.


Recently, reservoir computing (RC) has risen as an appealing approach due to its efficient training procedure~\cite{tanaka2019recent}.
RC is a particular type of recurrent neural network (RNN) that only contains a few numbers of trainable parameters and can be learned with a limited amount of training data.
Such properties enable its application in the online symbol detection task for both the OFDM system~\cite{zhou2019, xu2021rcstruct, xu2023DetectToLearn, li2023mmstructnet} and the OTFS system~\cite{zhou2022learningotfs}, where over-the-air (OTA) training data is extremely scarce.
Particularly, an RC-based approach for the OTFS system is first introduced in~\cite{zhou2022learningotfs} to perform the online subframe-based detection by adopting multiple RCs in the time domain.
This approach, however, directly applies the existing RC structure designed for the OFDM system in~\cite{zhou2019} and does not incorporate the domain knowledge of the OTFS system to reveal the full potential of RC.

In this work, we introduce a novel 2D-RC structure for the online subframe-based symbol detection task in the OTFS system.
The introduced 2D-RC retains the advantages of RC that can be learned with only a limited number of training pilots and has a low training complexity, which differentiates it from other NNs that require a large amount of training data and a long training time.
Furthermore, 2D-RC is novelly designed to have a 2D filtering structure tailored to equalizing the DD-domain 2D circular channel effect.
The 2D structure allows it to operate in the DD domain with a single NN, which is different from the previous RC-based approach in~\cite{zhou2022learningotfs} that requires multiple RCs in the time domain to track channel changes within one subframe.
As the number of RCs does not need to be configured, the 2D-RC can be more easily adopted in practice, making it a more general approach.
Simulation results show that 2D-RC can achieve substantial performance improvement over the previous multiple-RC approach under different modulation orders.
The results also demonstrate the advantages of 2D-RC over compared model-based schemes.

\textbf{Notations}: 
Non-bold letter, bold lowercase letter, bold uppercase letter, and bold Euler script letter, i.e., $x$, $\bm{x}$, $\bm{X}$, and $\mathbcal{X}$, denote scalar, vector, matrix, and tensor, respectively.
$\boldsymbol{F}_M$ is the normalized $M$-point discrete Fourier transform (DFT). 
$(\cdot)^\dag$ denotes the Moore–Penrose inverse.
$\langle\cdot\rangle_M$ is the modulo operator of divider $M$.
$\mathrm{vec}(\cdot)$ denotes the operation of vectoring the matrix by stacking along the columns, and $\mathrm{vec}^{-1}(\cdot)$ denotes the reverse operation.
$\odot$ denotes the matrix Hadamard product operation.
The $n$-mode Hadamard product between the matrix $\bm{U} \in \mathbb{C}^{I_n\times I_{n+1}}$ and the $N$-dimensional tensor $\mathbcal{X} \in \mathbb{C}^{I_1\times I_2 \times \dots \times I_N}$ is defined as
\begin{multline*}
    (\bm{U} \odot_n \mathbcal{X})[i_1, \dots,i_{n}, i_{n+1}, \dots, i_N] \\
    = U[i_{n}, i_{n+1}]\cdot X[i_1, \dots,i_{n}, i_{n+1}, \dots, i_N],
\end{multline*}
where $U[i_{n}, i_{n+1}]$ is the $(i_{n}, i_{n+1})$-th element in $\bm{U}$, and $X[i_1, \dots,i_{n}, i_{n+1}, \dots, i_N]$ is the $(i_1, \dots,i_{n}, i_{n+1}, \dots, i_N)$-th element in $\mathbcal{X}$.
The concatenation of two tensors $\mathbcal{X}_1$ and $\mathbcal{X}_2$ along the $n$-th dimension is represented by $\text{cat}_n(\mathbcal{X}_1, \mathbcal{X}_2)$.

\section{System Model}
\label{sec:sys_model}

\begin{figure}[t]
\centering
\includegraphics[width=0.6\linewidth]{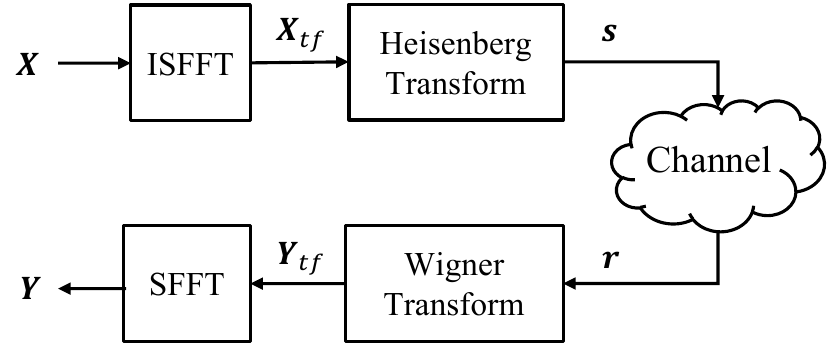}
\caption{OTFS system diagram.}
\label{fig:otfs_system}
\vspace{-1em}
\end{figure}

The transmitter and receiver structures in the OTFS system are shown in Fig.~\ref{fig:otfs_system}.
$Q$-th order quadrature amplitude modulation ($Q$-QAM) symbols are modulated in the DD domain, which forms the transmitted signal $\boldsymbol{X}$ of size $M\times N$.
$M$ and $N$ denote the number of delay bins and Doppler bins.
In this work, we consider the CP-OTFS system with practical rectangular transmit and received pulse shaping waveforms.
The CP-OTFS system can be implemented as an overlay of the OFDM system, where a cyclic prefix (CP) is added for each OFDM symbol in the subframe, i.e., $N$ CPs for one OTFS subframe.

\subsection{OTFS Transmitter and Receiver}
The transmitted signal $\boldsymbol{X}$ in the DD domain is converted to the TF domain through inverse symplectic finite Fourier transform (ISFFT) operation, which can be written as
\begin{align}
    \boldsymbol{X}_{tf} = \mathrm{ISFFT}(\boldsymbol{X}) = \boldsymbol{F}_M\boldsymbol{X}\boldsymbol{F}_N^H,
\end{align}
where $\boldsymbol{X}_{tf}$ represent the TF domain signal.
The TF domain signal is then transformed to the time domain signal $\bm{S}\in \mathbb{C}^{M\times N}$ for transmission by the Heisenberg transform and appended with CP.
The transmitted time-domain signal after adding CP can be expressed as
\begin{align}
    \bm{S} = \bm{A}_{cp}\bm{F}_M^H\bm{X}_{tf} = \bm{A}_{cp}\bm{X}\bm{F}_N^H,
\end{align}
where $\bm{A}_{cp} = [\bm{G}_{cp}^T, \bm{I}_M^T]^T \in \mathbb{R}^{(N_{cp} + M)\times M}$ is the CP addition matrix, $\bm{G}_{cp}$ with a size of $N_{cp}\times M$ consists of the last $N_{cp}$ rows of the identity matrix $\bm{I}_M$, and $N_{cp}$ is the CP length.
The vector form can be written as $\bm{s} = \mathrm{vec}(\bm{S}) \in \mathbb{C}^{MN\times 1}$.

Let $\bm{R} = \mathrm{vec}^{-1}(\boldsymbol{r}) \in \mathbb{C}^{M\times N}$ be the unfolded time domain received signal.
After removing CP, the received time domain signal $\bm{R}$ is converted back to the TF domain $\boldsymbol{Y}_{tf}$ through the Winger transform, which can be formulated by
\begin{align}
    \boldsymbol{Y}_{tf} = \bm{F}_M\bm{E}_{cp}\bm{R},
\end{align}
where $\bm{E}_{cp} = [\bm{0}_{M\times N_{cp}}, \bm{I}_M] \in \mathbb{R}^{M \times (N_{cp} + M)}$ is the CP removal matrix.
The DD domain received signal $\boldsymbol{Y}$ is obtained by applying the SFFT to the $\boldsymbol{Y}_{tf}$, which is expressed as
\begin{align}
    \boldsymbol{Y} = \mathrm{SFFT}(\boldsymbol{Y}_{tf}) = \boldsymbol{F}_M^H\boldsymbol{Y}_{tf}\boldsymbol{F}_N.
\end{align}

\subsection{Channel}
The channel response of the time-varying channel in the DD domain can be represented by
\begin{align*}
    h(\tau, \nu) = \sum_{i=0}^{P-1} h_i \delta(\tau - \tau_i)\delta(\nu - \nu_i),
\end{align*}
where $h_i$, $\tau_i$, and $\nu_i$ represent the complex path gain, delay, and Doppler shift of the $i$-th path; $P$ is the number of propagation paths.
The normalized delay shift $\mathcal{l}_i$ and Doppler shift $\kappa_i$ are given by $\tau_i = \frac{\mathcal{l}_i}{M\Delta f}$ and $\nu_i = \frac{\kappa_i}{T_f}$,
where $\mathcal{l}_i$ and $\kappa_i$ are not necessarily integers, $\Delta f$ is the subcarrier spacing, and $T_f$ is the total subframe duration.
In the time domain, the received signal can be expressed as~\cite{hadani2017orthogonal}
\begin{align*}
    r(t) = \int \int h(\tau, \nu)s(t-\tau)e^{j2\pi \nu (t-\tau)}d\tau d\nu + w(t),
\end{align*}
where $s(t)$ denotes the transmitted signal, and $w(t)$ is the additive Gaussian noise.

\subsection{Input-Output Relationship}
\label{sec:variants_otfs_system_io}
The DD-domain input-output relationship in the CP-OTFS system can be expressed as~\cite{das2020time}
\begin{equation}
\label{eq:cp_otfs_fractional}
\begin{split}
Y[l,k]
&=\sum_{i=0}^{P-1}\sum_{l'=0}^{M-1}\sum_{k'=0}^{N-1}h_i{z}^{k_p(N_{cp}+l-\mathcal{l}_i)}\psi_M(l' - \mathcal{l}_i)\\
&\times \psi_N(\kappa_i-k')X[\langle l-l'\rangle_N, \langle k-k'\rangle_M],
\end{split}
\end{equation}
where $Y[l,k]$ represents the ($l, k$)-th element of $\bm{Y}$, ${z} \triangleq e^{j\frac{2\pi}{(M+N_{cp})N}}$, and $\psi_M(x) \triangleq \frac{1}{M}e^{j\pi\frac{M-1}{M}x}\frac{\sin \pi  x }{\sin \pi  x/M }$.
For simplicity, we omit the noise term.
When $\mathcal{l}_i$ and $\kappa_i$ are integers, the~\eqref{eq:cp_otfs_fractional} simplifies to
\begin{align}
    \label{eq:cp_otfs_relationship}
    {Y}[l, k] = \sum_{i=0}^{P-1} h_i {z}^{k_i(N_{cp} + l-l_i)} {X}[\langle l-l_i\rangle_M, \langle k-k_i\rangle_N],
\end{align}
where $l_i$ and $k_i$ are the integer delay and integer Doppler.

\subsection{Problem formulation}


\begin{figure}
\centering%
\subfloat[Blockwise pilot pattern]{\label{a}\includegraphics[width=0.4\linewidth]{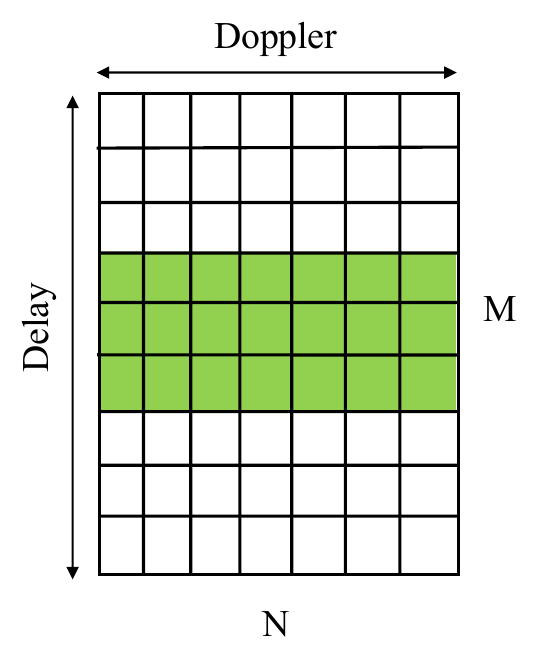}}%
\hspace{1em}
\subfloat[Spike pilot pattern]{\label{b}\includegraphics[width=0.4\linewidth]{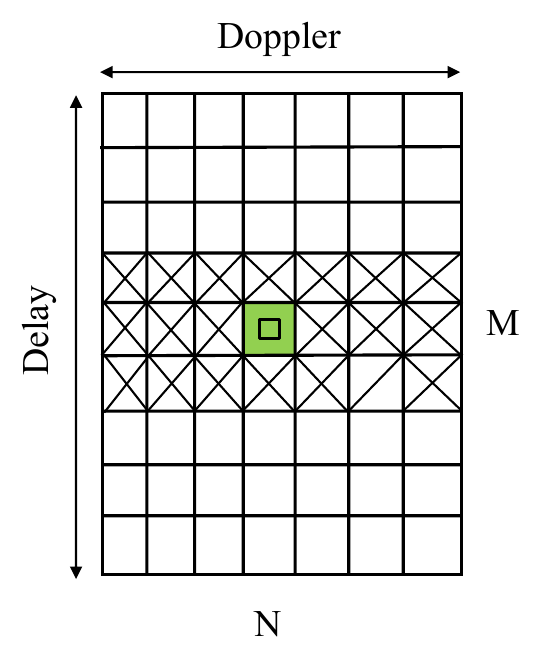}}
\caption{Pilot patterns.
The green grids are filled with known pilot symbols. 
The green grid with a square marker denotes the spike pilot.
The cross markers are guard symbols.
The blank region represents data symbol positions.
\label{figs:pilot_pattern}
}
\vspace{-1em}
\end{figure}

The OTFS symbol detection task is to recover the transmitted DD-domain symbol $\bm{X}$ in one OTFS subframe from the received signal.
To aid the detection of the unknown data symbols, the pilot symbols, which are known at both the transmitter and receiver sides, are inserted in each subframe.
In this paper, we consider two pilot structures: the blockwise pilot pattern and the spike pilot pattern, which are shown in Fig.~\ref{figs:pilot_pattern}.
Specifically, the pilot symbols are placed in a block of the subframe in the blockwise pilot structure.
In the spike pilot structure, a spike pilot is transmitted along with guard symbols surrounding it.
For learning-based approaches, the blockwise pilot pattern is adopted to learn the interference between the data symbols and pilot symbols. 
For model-based schemes that require knowledge of the CSI, the spike pilot pattern is utilized for channel estimation~\cite{raviteja2019embedded}.
More details about the choice of pilot patterns for learning-based and model-based approaches are provided in~\cite{zhou2022learningotfs}.

Denote $\bm{\Omega}$ as the pilot position indication matrix with $1$ indicating the pilot positions and $0$ otherwise.
For the introduced learning-based approach, the input to the NN is the received DD-domain signal $\bm{Y}$.
The training target is composed of the pilot symbols modulated in the DD domain, which can be denoted as $\bm{X}_{\mathrm{train}} \triangleq \bm{\Omega} \odot \bm{X}$.


\section{Introduced Approach -- 2D-RC}
\label{sec:introduced_approach}
In this section, we introduce the 2D-RC approach for online subframe-based symbol detection in the OTFS system.
RC is a class of RNNs for processing temporal or sequential data.
It consists of an input layer, an RNN-based reservoir, and an output layer.
The characteristic feature of RC is that the input layer and the reservoir weights are fixed after random initialization and only the output layer is updated through a simple linear regression.
The introduced 2D-RC retains the same fast and simple training process as RC, enabling it to perform online symbol detection on a subframe basis.
Moreover, it is uniquely designed to facilitate online symbol detection tailored towards the OTFS system.
Specifically, the DD-domain channel works as a 2D circular operation over the transmitted signal in the OTFS system as shown in~\eqref{eq:cp_otfs_fractional}.
To equalize this 2D circular channel effect, 2D-RC is designed to have a circular padding procedure and a 2D filtering process.
As 2D-RC works in the DD domain, it only requires a single NN for detection, as opposed to previous work~\cite{zhou2022learningotfs} that exploits multiple single-dimensional RCs (1D-RCs) to track the channel variations in the time domain.
For the remainder of this paper, we refer to the RC structure in~\cite{zhou2022learningotfs} as ``1D-RC".

\begin{figure*}
\centering
\includegraphics[width=0.6\linewidth]{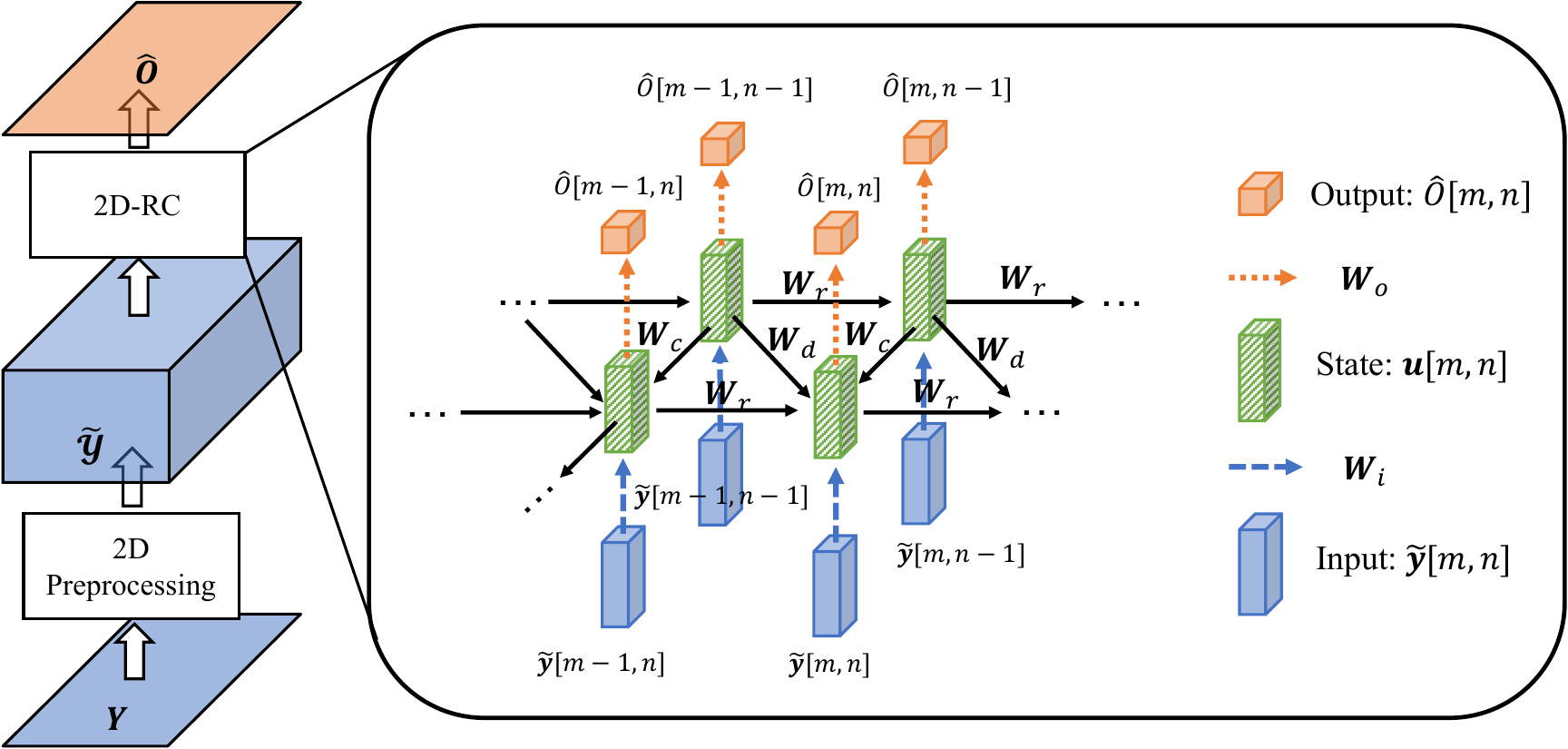}
\caption{2D-RC Structure. For simplicity, the nonlinear function and the extended state are ignored here. 
}
\label{figs:2d_rc_structure}
\end{figure*}
\begin{figure}
\centering
\includegraphics[width=0.6\linewidth]{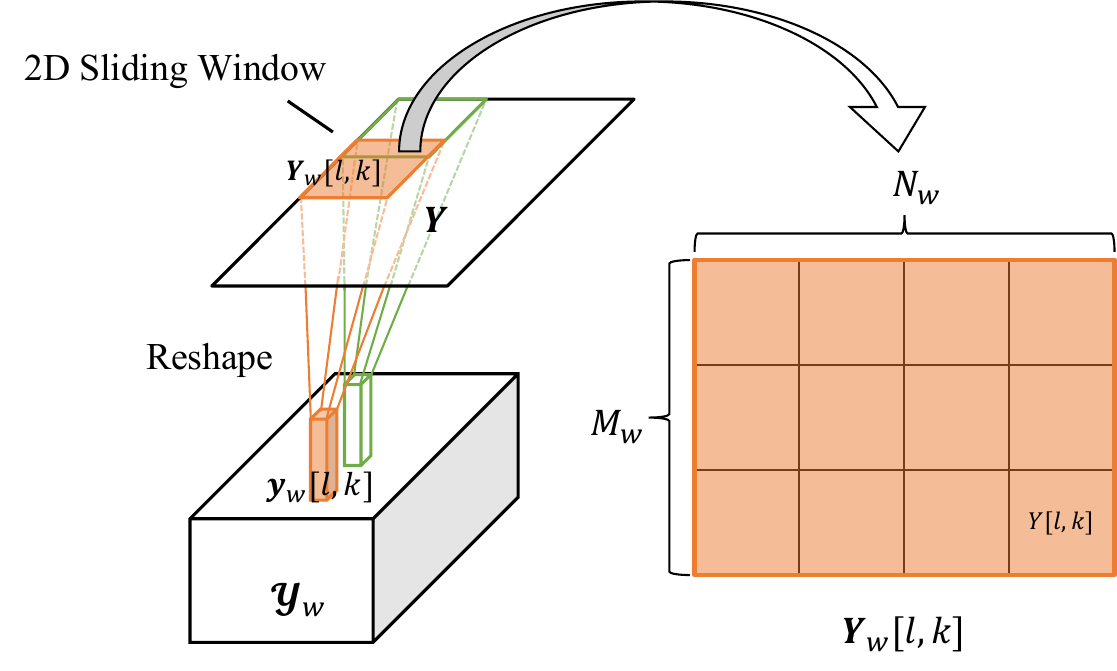}
\caption{The 2D windowing process in 2D-RC.}
\label{figs:2d_rc_pre_process}
\vspace{-1em}
\end{figure}

\subsection{Pre-processing}
The input to 2D-RC is the DD-domain received signal $\boldsymbol{Y} \in \mathbb{C}^{M\times N}$.
Two pre-processing procedures, including windowing and padding, are employed before the processing of 2D-RC.

\subsubsection{2D Windowing}
As discussed in~\cite{zhou2019}, adopting a sliding window to the input can increase the short-term memory of RC.
Following the same philosophy, we adopt the 2D windowing process, where a 2D sliding window with size $M_w \times N_w$ is employed to process the input.
The $M_w$ is the window size along the delay dimension and $N_w$ is the window size along the Doppler dimension.
For each ${Y}_c[l, k]$, the windowing region is obtained by $\bm{Y}_w[l, k] = \boldsymbol{Y}_c[l-M_w+1:l, k-N_w+1:k] \in \mathbb{C}^{M_w \times N_w}$.
When the $l<M_w-1$ or $k<N_w-1$, zeros are filled in the windowing region to maintain the window size of $M_w \times N_w$.
The windowed input is formed by $\bm{y}_w[l, k] = \mathrm{vec}(\mathrm{rev}(\bm{Y}_w[l, k]^T)) \in \mathbb{C}^{N_i}$, where $\mathrm{rev}(\cdot)$ stands for reserving the values in the matrix along both dimensions, $\mathrm{vec}(\cdot)$ represents vectoring the matrix by stacking along the columns, and $N_i = M_wN_w$.
By collecting all the $\bm{y}_w[l, k]$, we obtain an input tensor $\mathbcal{Y}_w \in \mathbb{C}^{N_i \times M \times N}$.
Fig.~\ref{figs:2d_rc_pre_process} visualizes the 2D windowing process.


\subsubsection{2D Circular Padding}
RC requires a degree of forgetfulness to remove the impact from the random initialization of the internal state~\cite{lukovsevivcius2009reservoir}.
Therefore, 2D-RC also needs to learn the optimal forget length to eliminate the impact of the initial state.
The 2D circular padding process is designed to facilitate the learning process of the optimal forget length.
Let $M_f$ and $N_f$ be the maximum forget length along the delay and Doppler dimension, respectively.
The 2D padded input $\tilde{\mathbcal{Y}} \in \mathbb{C}^{N_i \times (M + M_f) \times (N + N_f)}$ is obtained by concatenating the $\mathbcal{Y}_w$ along the second and third dimensions as the following:
\begin{align*}
    \tilde{\mathbcal{Y}} = & \nonumber
    \text{cat}_2( \text{cat}_3(\mathbcal{Y}_w,  \;\;           \mathbcal{Y}_w[:, :, 0:N_f-1]),\nonumber\\
               & \text{cat}_3(\mathbcal{Y}_w[:, 0:M_f-1, :],\mathbcal{Y}_w[:, 0:M_f-1, 0:N_f-1] )).
\end{align*}
Specifically, the 2D circular padding exploits the values at the start to pad at the end of the corresponding dimension.
The circular channel operation inspires the utilization of circular padding to initialize the states of 2D-RC.




\subsection{Structure of 2D-RC}

Denote $\tilde{\bm{y}}[m, n]\in \mathbb{C}^{N_i}$ as the ($m, n$)-th element along the second and third dimensions of the pre-processed input $\tilde{\mathbcal{Y}}$.
We design the state transition equation for 2D-RC as
\begin{align}
    \label{eq:2d_rc_state_transition}
    \bm{u}[m, n] &= f(\bm{W}_i\;\tilde{\bm{y}}[m, n] + \bm{W}_r\;\bm{u}[m-1, n] \nonumber\\
    &+ \bm{W}_d\;\bm{u}[m-1, n-1] + \bm{W}_c\;\bm{u}[m, n-1]),
\end{align}
where $\bm{u}[m, n] \in \mathbb{C}^{N_n}$ represent state vector for the $(m, n)$-th input; $N_n$ stands for the number of neurons; $\bm{W}_i \in \mathbb{C}^{N_n \times N_i}$ is the input weight matrix; $N_i$ denote the input dimension; $\bm{W}_r \in \mathbb{C}^{N_n \times N_n}$, $\bm{W}_c \in \mathbb{C}^{N_n \times N_n}$, and $\bm{W}_d \in \mathbb{C}^{N_n \times N_n}$ denote the reservoir weights along the row, column, and diagonal directions, respectively; $f(\cdot)$ is the hyperbolic tangent nonlinear activation function.
The input weights and reservoir weights are all randomly initialized by sampling from a uniform distribution.
In line with the 1D-RC approach, all reservoir weights are configured to be sparse with spectral radii less than $1$. 
The initial states $\bm{u}[-1, n]$, $\bm{u}[m, -1]$, and $\bm{u}[-1, -1]$ are all initialized as zero vectors.
The estimated output is obtained through the following output equation:
\begin{align}
    \hat{O}[m, n] = \bm{W}_o\;\tilde{\bm{u}}[m, n],\label{eq:2d_rc_output}
\end{align}
where $\tilde{\bm{u}}[m, n]= [\;\tilde{\bm{y}}[m, n]^T, \bm{u}[m, n]^T ]^T \in \mathbb{C}^{N_n+N_i}$ is the extended state formed by concatenating the input and the state, $\bm{W}_o \in \mathbb{C}^{1\times (N_n+N_i)}$ stands for the output weights.
By collecting all the state vectors ${\bm{u}}[m, n]$, the extended state vectors $\tilde{\bm{u}}[m, n]$, and the estimated output $\hat{O}[m, n]$, we can obtain the state tensor ${\mathbcal{U}} \in \mathbb{C}^{N_n \times (M + M_f) \times (N + N_f)}$, the extended state tensor $\tilde{\mathbcal{U}} \in \mathbb{C}^{(N_n+N_i) \times (M + M_f) \times (N + N_f)}$ and the estimated output matrix $\hat{\bm{O}} \in \mathbb{C}^{(M + M_f) \times (N + N_f)}$.
The structure is shown in Fig.~\ref{figs:2d_rc_structure}.


\subsection{Learning Algorithm}
Like other variants of RC, the input and reservoir weights of 2D-RC are all fixed after random initialization.
Only the output weights are updated during training.
The training objective for 2D-RC can be written as
\begin{align}
\min_{m_f \in \mathcal{L}_{m}, n_f \in \mathcal{L}_{n}, \bm{W}_{o}} ||\mathrm{vec}(\bm{\Omega} \odot \hat{\bm{O}}_{m_f, n_f}) - \mathrm{vec}(\bm{X}_{\mathrm{train}}) ||_2^2,
\label{eq:2d_objective_v1}
\end{align}
where $\hat{\bm{O}}_{m_f, n_f} = \hat{\bm{O}}[m_f:m_f+M-1, n_f:n_f+N-1] \in \mathbb{C}^{M \times N}$ represents the truncated output, $m_f$ is a forget length in the delay forget length set $\mathcal{L}_{m}$ with $M_f$ as the maximum delay forget length, and $n_f$ is a forget length in the Doppler forget length set $\mathcal{L}_{n}$ with $N_f$ as the maximum Doppler forget length.
Let $\tilde{\mathbcal{U}}_{m_f, n_f} = \tilde{\mathbcal{U}}[:, m_f:m_f+M-1, n_f:n_f+N-1] \in \mathbb{C}^{(N_n+N_i) \times M \times N}$ be the truncated extended state.
The masked truncated extended state tensor is denoted as $\bar{\mathbcal{U}}_{m_f, n_f} = \bm{\Omega} \odot_{2} \tilde{\mathbcal{U}}_{m_f, n_f}$, where $\odot_{2}$ represents conducting the Hadamard product along the second and third dimensions.
The masked truncated extended state matrix $\bar{\bm{U}}_{m_f, n_f} = \mathrm{vec}_2(\bar{\mathbcal{U}}_{m_f, n_f}) \in \mathbb{C}^{(N_n+N_i) \times MN}$ is formed by vectoring the last two dimensions of $\bar{\mathbcal{U}}_{m_f, n_f}$ with $\mathrm{vec}_2(\cdot)$ denoting vectoring along the second and third dimensions.
Then by substituting \eqref{eq:2d_rc_output} into \eqref{eq:2d_objective_v1}, the learning objective function becomes
\begin{align}
    \min_{m_f \in \mathcal{L}_{m}, n_f \in \mathcal{L}_{n}, \bm{W}_{o}} ||\bm{W}_o\bar{\bm{U}}_{m_f, n_f} - (\mathrm{vec}(\bm{X}_{\mathrm{train}}))^T||_2^2.
\end{align}
The joint optimization problem is solved by alternatively minimizing the objective with the forget length and the output weights.
Specifically, we first fix the forget length $m_f$ and $n_f$ and obtain the trained output weights by the LS solution
\begin{align}
    \label{eq:2d_output_mtx_estimation}
    \hat{\bm{W}}_o^{(m_f, n_f)} =  (\mathrm{vec}(\bm{X}_{\mathrm{train}}))^T\bar{\bm{U}}_{m_f, n_f}^\dag.
\end{align}
Then the optimal forget lengths along the delay dimension and Doppler dimension are learned by finding the length that minimizes the loss with
\begin{align}
    \hat{m}_f, \hat{n}_f \!= \!\!\!\!\argmin_{m_f \in \mathcal{L}_{m}, n_f \in \mathcal{L}_{n}} \!\!||\hat{\bm{W}}_o^{(m_f, n_f)} \bar{\bm{U}}_{m_f, n_f} \!- \!(\mathrm{vec}(\bm{X}_{\mathrm{train}}))^T||_2^2.
\end{align}
Instead of searching through all the possible delay and Doppler forget length pairs, we first find the optimal Doppler forget length and then find the optimal delay forget length to reduce the training complexity.

\subsection{Testing with 2D-RC}
At the testing stage, the transmitted symbols $\hat{\bm{x}} \in \mathbb{C}^{1 \times MN}$ are estimated by
\begin{align}
    \label{eq:2d_rc_output_eq}
    \hat{\bm{x}} = \mathcal{Q}(\hat{\bm{W}}_o^{(\hat{m}_f, \hat{n}_f)} \;\tilde{\bm{U}}_{\hat{m}_f, \hat{n}_f}),
\end{align}
where $\hat{\bm{W}}_o^{(\hat{m}_f, \hat{n}_f)}$ is the trained output matrix when utilizing the forget length $\hat{m}_f$ and $\hat{n}_f$, $\tilde{\bm{U}}_{\hat{m}_f, \hat{n}_f} =\mathrm{vec}_2(\tilde{\mathbcal{U}}_{\hat{m}_f, \hat{n}_f})  \in \mathbb{C}^{(N_n+N_i) \times MN}$ is obtained by vectoring the truncated extended state tensor $\tilde{\mathbcal{U}}_{\hat{m}_f, \hat{n}_f}$ with forget length $\hat{m}_f$ and $\hat{n}_f$, and $\mathcal{Q}(\cdot)$ is the quantization operation that maps the output to the constellation points.
The transmitted data symbols are extracted with $\hat{\bm{X}}_{\mathrm{data}} = \bar{\bm{\Omega}} \odot \hat{\bm{X}}$, 
where $\hat{\bm{X}} =\mathrm{vec}^{-1}(\hat{\bm{x}}) \in \mathbb{C}^{M \times N}$ and $\bar{\bm{\Omega}}$ is the complement of $\bm{\Omega}$.




\section{Numerical Experiments}
\label{sec:experiments}



In this section, we evaluate the performance of 2D-RC for symbol detection in the OTFS system.
We set $N=14$ and $M=1024$ following the 3GPP 5G NR standard~\cite{std3gpp36211, std3gpp36212}.
The carrier frequency is $4$ GHz and subcarrier spacing is $15$ kHz.
The 3GPP 5G NR clustered delay line (CDL) channel with delay profile ``CDL-C"~\cite{std3gpp38901} is considered with a delay spread of $10$ ns and user velocity of $150$ km/h.
The pilot overhead is $4.69 \%$, which is set to satisfy the pilot overhead requirement specified in~\cite{std3gpp36211, std3gpp36212}.
All the compared approaches adopt the same training overhead for a fair comparison.
The parameters of 2D-RC are set as $N_n=6$, $M_w=4$, $N_w=14$, and $l_c=7$.
The delay forget length and Doppler forget length are searched in the range of $7$ to $8$ and the range of $13$ to $14$, respectively.
The spectral radii of all the reservoir weights are configured as $0.9$ and the sparsities are set as $0.6$.
The quantization operation is set as the minimum distance mapping.

\begin{figure}
\centering
\includegraphics[width=0.8\linewidth]{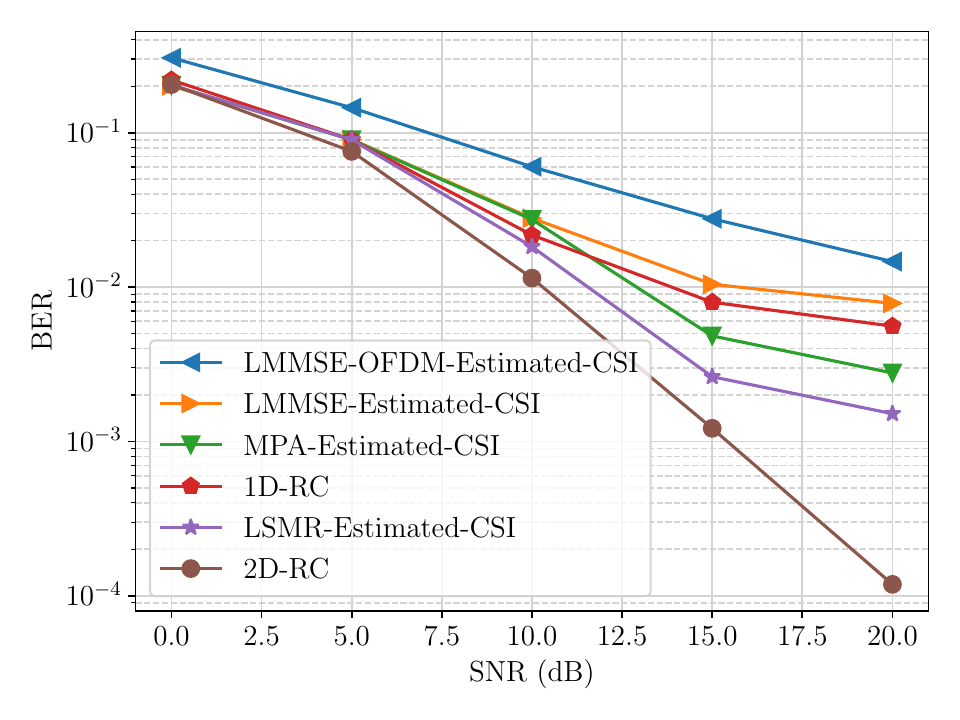}
\caption{BER comparison under QPSK.}
\label{figs:ber_cp_rec_pulse_qpsk_10ns}
\vspace{-1em}
\end{figure}

We compare the performance of 2D-RC with the 1D-RC~\cite{zhou2022learningotfs}, LMMSE detectors, least squares minimum residual (LSMR) based method~\cite{qu2021TCOMLowcomplexSD}, and MPA~\cite{raviteja2018interference}.
For the 1D-RC approach, the number of neurons is $12$, and $7$ 1D-RCs are adopted for detection.
The LMMSE detector in the OTFS system is implemented in the time domain with the block-wise channel inverse to reduce the computational complexity~\cite{hong2022delay}.
The LMMSE method in the OFDM system performs detection in the TF domain with the LMMSE estimated channel in the TF domain~\cite{hoeher1997two}.
For the iterative LSMR-based scheme, the number of iterations for interference cancellation is set as $5$ and $10$ for QPSK and $16$ QAM, respectively. 
The number of iterations for LSMR is $15$ for both modulation schemes.
For the MPA method, the number of iterations is $30$ and the damping factor is set as $0.6$.
All the estimated CSI in the OTFS system is obtained by the approach in~\cite{raviteja2019embedded} with the spike pilot pattern.
The power of the spike pilot is set to ensure that the OTFS subframe with the spike pilot pattern has approximately the same PAPR as utilizing the blockwise pilot pattern.
The reason is that a high peak-to-average power ratio (PAPR) may compel the power amplifier (PA) to operate in the non-linear region, resulting in signal distortion and spectral spreading~\cite{rahmatallah2013peak}.
Depending on the tested SNR and modulation order, the received pilot signal-to-noise ratio (SNR) ranges from around $20$ dB to $47$ dB, which covers the commonly considered pilot SNRs in existing works, e.g.,~\cite{zhou2022learningotfs, liu2021message}.

\begin{figure}
\centering
\includegraphics[width=0.8\linewidth]{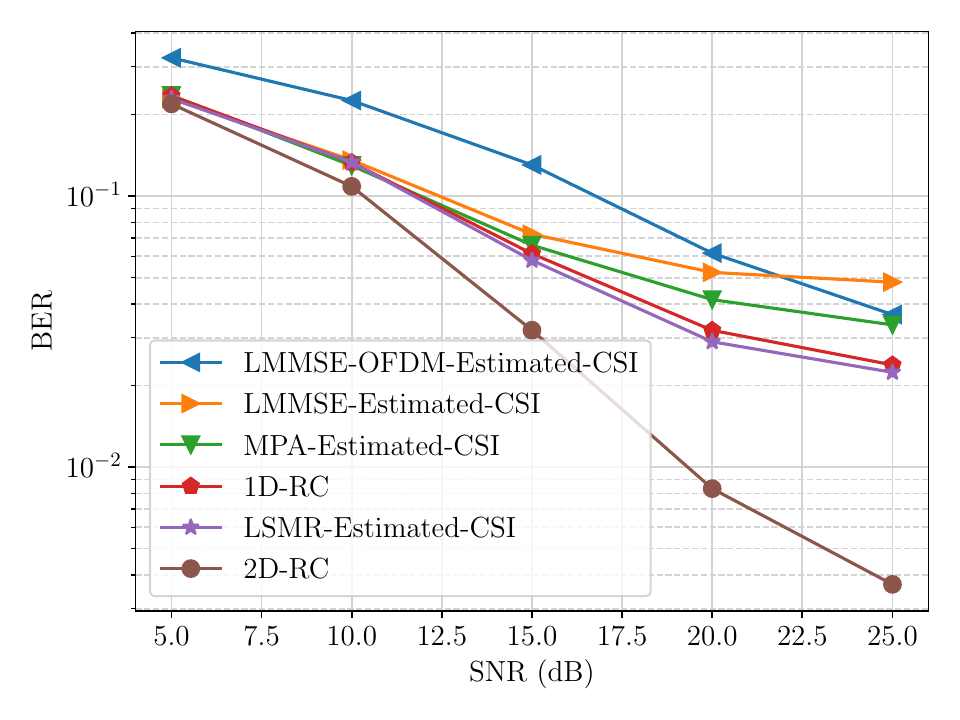}
\caption{BER comparison under 16 QAM.}
\label{figs:ber_cp_rec_pulse_16qam_10ns}
\vspace{-1em}
\end{figure}

In Fig.~\ref{figs:ber_cp_rec_pulse_qpsk_10ns} and Fig.~\ref{figs:ber_cp_rec_pulse_16qam_10ns}, we compare the bit error rate (BER) performance of different approaches in the CP-OTFS system under the QPSK and 16 QAM modulations, respectively.
Compared with the existing learning-based 1D-RC method, 2D-RC is demonstrated to have better performance under both the QPSK and 16 QAM modulations, especially in the high SNR regime.
Note that $7$ RCs are utilized in the 1D-RC approach, while only a single NN is exploited for 2D-RC.
The reason is that the 1D-RC method directly adopts the existing RC architecture in the time domain and does not leverage domain knowledge of the OTFS system for its design.
When operating in the time domain, multiple RCs are required to track the changes in the time-varying channel.
Instead, 2D-RC incorporates the 2D circular structure in the DD-domain input-output relationship into its design.
By incorporating structural knowledge, even with a single NN, 2D-RC is more effective than the 1D-RC method that adopts multiple RCs.
The 2D-RC also outperforms compared model-based approaches, i.e., LMMSE, MPA, and the LSMR-based approach, when employing the estimated channel.
Different from the model-based approaches that rely on the knowledge of CSI, the introduced learning-based 2D-RC approach does not require channel knowledge.
Therefore, the performance of 2D-RC is not affected by the accuracy of channel estimates and can be more easily adopted in practical scenarios when it is hard to obtain an accurate CSI.
Furthermore, all the considered OTFS-based detectors in the CP-OTFS system are shown to perform better than the LMMSE approach in the OFDM system in mid to low SNR regimes.

\section{Conclusion}
\label{sec:conclusion}

In this paper, we introduce a learning-based 2D-RC approach for the symbol detection task in the OTFS system.
As with other variants of RC, the introduced 2D-RC approach can conduct online subframe-based symbol detection with a limited amount of training data.
Extending beyond our previous RC-based approach, the 2D-RC is based on a novel structure tailored to equalizing the 2D circular channel effect in the DD domain of the OTFS system.
By employing the 2D structure, the 2D-RC approach is able to perform online subframe-based detection in the DD domain with a single NN.
Simulation results demonstrate that 2D-RC significantly outperforms the previous 1D-RC approach that requires multiple RCs to track the channel changes across time.
Furthermore, compared with the model-based approaches, 2D-RC does not require any channel knowledge or system assumptions.
It is shown to have better performance than the LMMSE and MPA with the estimated CSI across different modulation orders.




\ifCLASSOPTIONcaptionsoff
  \newpage
\fi



\bibliographystyle{IEEEtran}

\bibliography{IEEEabrv,ref.bib}

\begin{thebibliography}{10}
\providecommand{\url}[1]{#1}
\csname url@samestyle\endcsname
\providecommand{\newblock}{\relax}
\providecommand{\bibinfo}[2]{#2}
\providecommand{\BIBentrySTDinterwordspacing}{\spaceskip=0pt\relax}
\providecommand{\BIBentryALTinterwordstretchfactor}{4}
\providecommand{\BIBentryALTinterwordspacing}{\spaceskip=\fontdimen2\font plus
\BIBentryALTinterwordstretchfactor\fontdimen3\font minus
  \fontdimen4\font\relax}
\providecommand{\BIBforeignlanguage}[2]{{%
\expandafter\ifx\csname l@#1\endcsname\relax
\typeout{** WARNING: IEEEtran.bst: No hyphenation pattern has been}%
\typeout{** loaded for the language `#1'. Using the pattern for}%
\typeout{** the default language instead.}%
\else
\language=\csname l@#1\endcsname
\fi
#2}}
\providecommand{\BIBdecl}{\relax}
\BIBdecl

\bibitem{series2015imt}
M.~Series, ``{IMT} {Vision--Framework} and overall objectives of the future
  development of {IMT} for 2020 and beyond,'' \emph{Recommendation ITU}, vol.
  2083, no.~0, 2015.

\bibitem{hadani2017orthogonal}
R.~Hadani, S.~Rakib, M.~Tsatsanis, A.~Monk, A.~J. Goldsmith, A.~F. Molisch, and
  R.~Calderbank, ``Orthogonal time frequency space modulation,'' in \emph{2017
  IEEE Wireless Commun. Netw. Conf.}, 2017, pp. 1--6.

\bibitem{zou2021low}
T.~Zou, W.~Xu, H.~Gao, Z.~Bie, Z.~Feng, and Z.~Ding, ``Low-complexity linear
  equalization for {OTFS} systems with rectangular waveforms,'' in \emph{2021
  IEEE Intl. Conf. on Commun. (ICC)}.\hskip 1em plus 0.5em minus 0.4em\relax
  IEEE, 2021, pp. 1--6.

\bibitem{raviteja2018interference}
P.~Raviteja, K.~T. Phan, Y.~Hong, and E.~Viterbo, ``Interference cancellation
  and iterative detection for orthogonal time frequency space modulation,''
  \emph{{IEEE} Trans. Wireless Commun.}, vol.~17, no.~10, pp. 6501--6515, 2018.

\bibitem{liu2021message}
F.~Liu, Z.~Yuan, Q.~Guo, Z.~Wang, and P.~Sun, ``Message passing-based
  structured sparse signal recovery for estimation of {OTFS} channels with
  fractional {Doppler} shifts,'' \emph{{IEEE} Trans. Wireless Commun.},
  vol.~20, no.~12, pp. 7773--7785, 2021.

\bibitem{zhang2021low}
H.~Zhang and T.~Zhang, ``A low-complexity message passing detector for {OTFS}
  modulation with probability clipping,'' \emph{{IEEE} Wireless Commun. Lett.},
  vol.~10, no.~6, pp. 1271--1275, 2021.

\bibitem{enku2021two}
Y.~K. Enku, B.~Bai, F.~Wan, C.~U. Guyo, I.~N. Tiba, C.~Zhang, and S.~Li,
  ``Two-dimensional convolutional neural network-based signal detection for
  {OTFS} systems,'' \emph{{IEEE} Wireless Commun. Lett.}, vol.~10, no.~11, pp.
  2514--2518, 2021.

\bibitem{zhang2022gaussian}
X.~Zhang, L.~Xiao, S.~Li, Q.~Yuan, L.~Xiang, and T.~Jiang, ``Gaussian {AMP}
  aided model-driven learning for {OTFS} system,'' \emph{{IEEE} Commun. Lett.},
  vol.~26, no.~12, pp. 2949--2953, 2022.

\bibitem{tanaka2019recent}
G.~Tanaka, T.~Yamane, J.~B. H{\'e}roux, R.~Nakane, N.~Kanazawa, S.~Takeda,
  H.~Numata, D.~Nakano, and A.~Hirose, ``Recent advances in physical reservoir
  computing: A review,'' \emph{Neural Netw.}, vol. 115, pp. 100--123, 2019.

\bibitem{zhou2019}
Z.~Zhou, L.~Liu, and H.-H. Chang, ``Learning for detection: {MIMO-OFDM} symbol
  detection through downlink pilots,'' \emph{{IEEE} Trans. Wireless Commun.},
  vol.~19, no.~6, pp. 3712--3726, 2020.

\bibitem{xu2021rcstruct}
J.~Xu, Z.~Zhou, L.~Li, L.~Zheng, and L.~Liu, ``{RC-Struct}: A structure-based
  neural network approach for {MIMO-OFDM} detection,'' \emph{{IEEE} Trans.
  Wireless Commun.}, vol.~21, no.~9, pp. 7181--7193, 2022.

\bibitem{xu2023DetectToLearn}
J.~Xu, L.~Li, L.~Zheng, and L.~Liu, ``Detect to learn: Structure learning with
  attention and decision feedback for {MIMO-OFDM} receive processing,''
  \emph{{IEEE} Trans. Commun.}, pp. 1--1, 2023.

\bibitem{li2023mmstructnet}
L.~Li, J.~Xu, L.~Zheng, and L.~Liu, ``Real-time machine learning for multi-user
  massive {MIMO}: Symbol detection using {Multi-Mode} {StructNet},''
  \emph{{IEEE} Trans. Wireless Commun.}, pp. 1--1, 2023.

\bibitem{zhou2022learningotfs}
Z.~Zhou, L.~Liu, J.~Xu, and R.~Calderbank, ``Learning to equalize {OTFS},''
  \emph{{IEEE} Trans. Wireless Commun.}, vol.~21, no.~9, pp. 7723--7736, 2022.

\bibitem{das2020time}
S.~S. Das, V.~Rangamgari, S.~Tiwari, and S.~C. Mondal, ``Time domain channel
  estimation and equalization of {CP-OTFS} under multiple fractional dopplers
  and residual synchronization errors,'' \emph{IEEE Access}, vol.~9, pp.
  10\,561--10\,576, 2020.

\bibitem{raviteja2019embedded}
P.~Raviteja, K.~T. Phan, and Y.~Hong, ``Embedded pilot-aided channel estimation
  for {OTFS} in delay--{Doppler} channels,'' \emph{{IEEE} Trans. Veh.
  Technol.}, vol.~68, no.~5, pp. 4906--4917, 2019.

\bibitem{lukovsevivcius2009reservoir}
M.~Luko{\v{s}}evi{\v{c}}ius and H.~Jaeger, ``Reservoir computing approaches to
  recurrent neural network training,'' \emph{Comput. Sci. Review}, vol.~3,
  no.~3, pp. 127--149, 2009.

\bibitem{std3gpp36211}
\emph{5G; NR; Physical channels and modulation}, 3GPP Std. TS 36.211, Rev.
  16.2.0, 2020.

\bibitem{std3gpp36212}
\emph{5G; NR; Multiplexing and channel coding}, 3GPP Std. TS 36.212, Rev.
  16.2.0, 2020.

\bibitem{std3gpp38901}
\emph{5G; Study on channel model for frequencies from 0.5 to 100 {GHz}}, 3GPP
  Std. TR 38.901, Rev. 16.1.0, 2020.

\bibitem{qu2021TCOMLowcomplexSD}
H.~Qu, G.~Liu, L.~Zhang, S.~Wen, and M.~A. Imran, ``Low-complexity symbol
  detection and interference cancellation for {OTFS} system,'' \emph{{IEEE}
  Trans. Commun.}, vol.~69, no.~3, pp. 1524--1537, 2021.

\bibitem{hong2022delay}
Y.~Hong, T.~Thaj, and E.~Viterbo, \emph{Delay-{Doppler} Communications:
  Principles and Applications}.\hskip 1em plus 0.5em minus 0.4em\relax Academic
  Press, 2022.

\bibitem{hoeher1997two}
P.~Hoeher, S.~Kaiser, and P.~Robertson, ``Two-dimensional pilot-symbol-aided
  channel estimation by wiener filtering,'' in \emph{1997 IEEE int. conf. on
  acoust., speech, and signal process.}, vol.~3, pp. 1845--1848.

\bibitem{rahmatallah2013peak}
Y.~Rahmatallah and S.~Mohan, ``Peak-to-average power ratio reduction in {OFDM}
  systems: A survey and taxonomy,'' \emph{{IEEE} Commun. Surveys Tuts.},
  vol.~15, no.~4, pp. 1567--1592, 2013.

\end{thebibliography}
\end{document}